\documentclass[twocolumn,amsmath,amssymb,longbibliography,superscriptaddress]{revtex4-1}
\usepackage{graphicx}
\usepackage{dcolumn}
\usepackage{bm}
\usepackage{subcaption}
\usepackage{multirow}
\usepackage{ragged2e}
\usepackage[usenames,dvipsnames]{xcolor}

\usepackage{adjustbox}
\usepackage{xcolor}
\usepackage{soul}

\begin{document}
\justifying

\preprint{ \color{Green} DRAFT v1.0 - \today}

\title{\Large Multiple droplets dynamics on cylindrical fiber}

\author{J. Van Hulle}
\affiliation{GRASP, Institute of Physics B5a, University of Li\`ege, B4000 Li\`ege, Belgium}
\affiliation{Dynamics of Fluids, Department of Experimental Physics, Saarland University, 66123 Saarbr{\"u}cken, Germany }
\author{N. Vandewalle}
\affiliation{GRASP, Institute of Physics B5a, University of Li\`ege, B4000 Li\`ege, Belgium}
\author{Z. Pan}
\affiliation{Pan-Lab, University of Waterloo, 200 University Avenue West, Waterloo, Ontario, Canada N2L 3G1}


{\begin{abstract}
We investigate the descent of a single droplet on a vertical fiber and its role in generating additional droplets. As the droplet slides, it deposits a non-uniform liquid film whose thickness decreases along the fiber. The thicker portion near the release point undergoes a Rayleigh-Plateau instability, leading to the formation of a new droplet. This newly formed droplet initially accelerates by accumulating fluid from the film ahead, then shifts to a decelerating phase once fluid loss at its rear dominates. Ultimately, it merges with the original droplet. This refeeding allows the droplet to travel further. 
Successive generations of droplets can similarly arise, governed by the evolving film thickness. We develop theoretical models that capture each phase of motion and discuss the experimental and theoretical aspects of this cascading cycle of droplet generation. 
 \end{abstract}
}

\vskip 2 mm
\hrule 
\vskip -3 mm

\maketitle
\section{Introduction}

Liquid films are a broad research area encompassing both natural systems and technological processes.
They span length scales from micron-thick tear film on the cornea to meter-thick films in lava flows~\cite{Wong1996,  Griffiths2000}. 
Such films show diverse behaviors.
For example, window panels or gutters under heavy rainfall display quasiperiodic wave patterns, paint applied on walls may exhibit fingering, and liquid coatings on fibers may destabilize into droplet configurations.

Classical coating theory, developed by Landau, Levich, and Derjaguin, describes how a thin film is entrained on a flat plate withdrawn from a liquid bath. 
The resulting film thickness depends on a balance between viscous drag and capillary suction and scales with the withdrawal speed~\cite{Landau1988}. 
When the coated flat substrate is tilted, the film flows under gravity, which leads to diverse phenomena. 
The dynamics is mainly a results of the competition between viscous, inertial, and interfacial effects.  
When inertia is negligible, perturbation at the contact line can grow, forming triangular sawtooth patterns or straight-edged finger-like rivulets~\cite{Huppert1982, Hocking1999, Johnson1999}.
However, when inertial effects become significant, disturbances from the noise at the inlet lead to wave patterns, either short-wavelength sinusoidal shapes or pulse-like solitary waves~\cite{Park2003, Nosoko2004, Scheid2006}.
Increasing the flow rate promotes chaotic wave behavior.



Replacing the flat plate with a cylindrical fibre removes one transverse dimension and introduces a curvature at the interface. 
Coating theory is similar to the case of the flat substrate, but the film thickness additionally depends on the fiber diameter~\cite{Quere1999}.
Film flow dynamics on vertical fibers results in various patterns, including wave formation, pulses, and droplet formation through Rayleigh-Plateau instability~\cite{Kalliadasis1994, Craster2006, Gabbard2021PRF}.
Observed behaviors depend on fiber sizes, flow rate, and nozzle diameter~\cite{Smolka2008, Kliakhandler2001, Sadeghpour2017, Gabbard2021ExpFluids, Guo2022}. 
When increasing the flow rate or fiber diameter, the system transitions from evenly spaced droplets cyclically coalescing with smaller beads to a steady Rayleigh-Plateau regime with periodic bead trains, and finally to an unsteady regime with random coalescence and irregular patterns.
When inertia dominates surface tension effects, capillary wave trains form along the liquid film~\cite{Duprat2009}.
Quéré further provided a cutoff film thickness below which vertical coatings remain cylindrically stable~\cite{Quere1990EPL}. 

Beyond film flows, interactions between individual droplets with flat or curved substrates have also been widely studied. Droplets spread on flat substrates or adopt spherical shapes depending on surface wettability and substructures~\cite{Bonn2009, Quere2002}. 
Conversely, on fibers, droplets depict two distinct configurations based on fiber size and liquid properties~\cite{Carroll1976, McHale2002}.
Small droplets adopt asymmetrical clamshell configurations, while larger droplets adopt symmetric barrel geometry around the fiber.
Furthermore, the dynamics of droplet sliding under gravity on both flat and curved substrates have been extensively described for various geometrical cases~\cite{Weyer2017, Khattak2024, Leonard2023, Kern2024, VanHulle2024} and wetting conditions~\cite{Furmidge1962, Mahadevan1999, Gilet2010}.


Yet, some mysteries are still to be unraveled. In particular, Quéré \textit{et al.} highlighted in~\cite{Quere1990Science} predator/prey-like interactions between droplets on a vertical fiber resulting from the Rayleigh-Plateau instability: ``When [...] the film is thick [...], drops appear quickly and fall along the fiber. 
If one drop is slightly larger than the other, it falls faster and in passing swallows the rest". 
To the knowledge of the authors, this phenomenon remains understudied despite its common occurrence in fiber- and droplet-based experiments.

This study investigates the dynamics of a single droplet descending along a vertical dry fiber, including phenomena induced by its passage. 
As the droplet descends, it coats the dry fiber with a thin film of non-uniform thickness while gradually decelerating due to its decreasing weight.
The deposited film may rapidly bead up, forming droplets that descend, catch up to their predecessors, and leave behind their own coating films. 
This iterative cascade of droplets on a fiber sits in between coating processes, film dynamics on fibers, and droplet behaviors.

Throughout this study, we use terms such as ``mother'' and ``daughter'' droplets, and the ``generation'' of a droplet family, to describe the recurring pattern of droplet dynamics on a fiber. 
This choice is partly inspired by an intriguing parallel to Gabriel García Márquez's novel One Hundred Years of Solitude, in which the Buendía family is doomed to repeat its fate, shaped by inevitable forces beyond their control~\cite{GarciaMarquez}. 
Similarly, in our experiments, each generation of the droplets inherits the wetting conditions left by its progenitors and sets the wetting for descendants.
The repeated birth, motion, and eventual merging or perishing of droplets create a cycle, governed by perpetual physical laws that mirror the generational patterns described in the novel.

\begin{figure*}[ht]
\centering
\centering
\includegraphics[width=\textwidth]{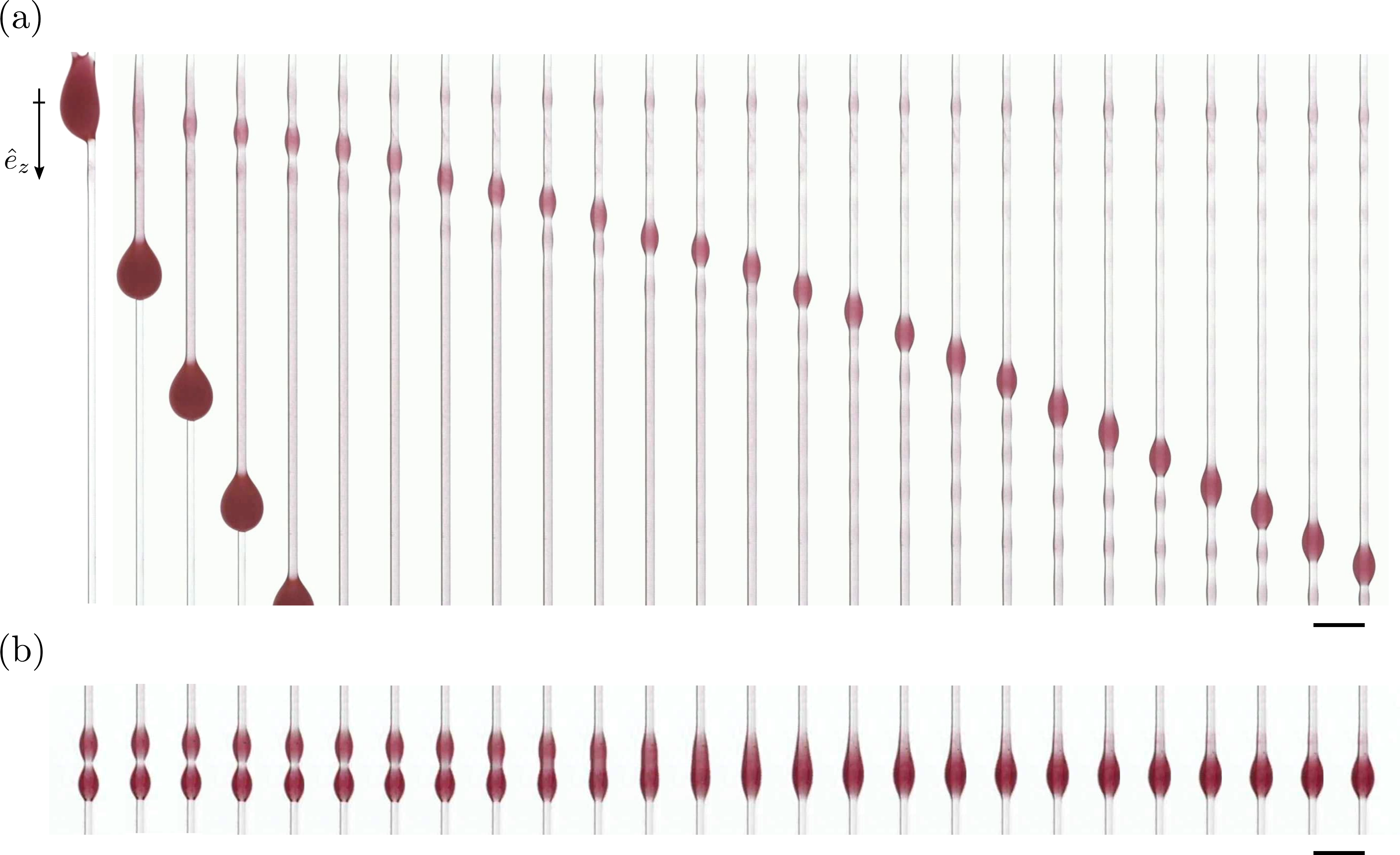}
\caption{\justifying (a) Red dyed silicone oil droplet descending a vertical nylon fiber. The time interval between each image is $0.68$~s. As the droplet slides down, a liquid film is left behind. The thickness of this liquid film is larger when the droplet speed is higher. Therefore, at the release point of the initial droplet, the so-called mother or generation $0$ droplet, the film is the thickest. The Rayleigh-Plateau instability takes place, which creates a tiny new droplet, termed daughter or generation $1$ droplet. This daughter droplet slides on the film left by the mother droplet. (b) Same experiment but down the fiber. The mother droplet is positioned below, while the daughter droplet follows above. The time interval between each image is $0.017$~s. One observes that the daughter droplet finally catches the mother droplet, leading to a merging of both droplets. The experimental data for (a) and (b) are the same, $\Omega = 5~\mu$L, $\nu = 100$~cSt, and $d = 0.3$~mm. The scale bars correspond to $2$~mm.}
\label{fig_merging}
\end{figure*}

\section{Methods}
We used a pipette (Eppendorf Xplorer) to release a droplet on a vertical fiber made of nylon (fishing thread). 
These fibers have a circular cross-section with diameters ranging from $d = 0.2$~mm to $d = 0.45$~mm. 
The liquid used is silicone oil with kinematic viscosities of $\nu = 50$~cSt and $100$ cSt. 
The tested volumes were $\Omega = 3, 5,$ and $7~\mu$L. 
The silicone oil is dyed red, and a white lighting source is positioned behind the fiber to enhance contrast. 
Before each experiment, the fiber was cleaned with isopropanol and distilled water and left to dry. 
Each experiment was repeated five times. 

The motion of the droplets was recorded with a CCD camera positioned in front of the fiber. 
The camera setup allows a field of view of $13$~cm while still capturing the millimeter-scale droplets at $35$~fps. 
The positions of the droplets $z$ over time $t$ were extracted using a Python script from calibrated video images. 
For imaging the experiment, closer views were recorded at $60$~fps as in Fig.~\ref{fig_merging} (a) and (b).

\section{Results}
When the first droplet is deposited on a vertical dry fiber, it slides and coats the fiber with a liquid film (see Fig.~\ref{fig_merging} (a)), and we call this droplet the mother droplet or the `generation~0'. 
Meanwhile, as the mother droplet loses volume when it coats the fiber, it decelerates.
The thickness of the coating film left behind by a droplet is determined by its speed: the faster the droplet moves, the thicker the film, following the Landau-Levich-Derjaguin theory \cite{Landau1988, Quere1999}.
Consequently, the thickness of the coating film created by the mother droplet reduces along the fiber, with the thickest section located near the initial deposition point.

A liquid film coated on a fiber tends to minimize its surface area and bead up due to Rayleigh-Plateau instability. 
The growth rate of the droplet is proportional to the inverse of the film thickness, and the droplet size is proportional to the film thickness.
Since the film is thickest at the initial release point, a new droplet beads up rapidly there. 
We call this droplet the daughter droplet, which is the first and the biggest droplet in `generation~1'. 
This new droplet travels along the wetted fiber, following the non-uniform liquid film left by its mother. 
Along its path, it engulfs other droplets of the same generation, whose slower growth results from the thinner liquid film.
As the daughter droplet descends, a liquid coating is again left behind, potentially enabling the formation of the next generation of droplets.
The subsequent droplet that emerged at the initial deposition by the daughter's residual film is termed `granddaughter' droplet or generation 2.
This process may repeat and produce a `great-granddaughter' droplet (generation 3), and so on.

A distinction must be made between generation 0 and the following generations, which results in distinct dynamics for generations 0 and generation-$k$ $(k\geq 1)$. 
Generation 0 slides on a dry fiber, while generations 1 and beyond slide on a wetted fiber. 
Notably, it is observed that both generations merge after traveling a certain distance, as shown in Fig. \ref{fig_merging} (b). 
The mother and the daughter droplets are destined to meet again, and it is only a matter of time before it occurs. 
The merging refeeds the mother droplet, enabling it to continue its descent.

In our experiments, the motion of the droplet is affected by gravity as revealed when calculating the Bond number Bo $= \rho g \Omega/2 \pi \gamma d$, where $\rho$ is the fluid density, $g$ is the gravitational acceleration, and $\gamma$ is the surface tension.
Taking experimentally tested values of $\Omega$ and $d$ yields Bo $\in [0.49, 2.55]$, indicating that both gravity and surface tension influence the droplet's motion .
The range of applicable Bo to our case study is further discussed in section I.V. D. 
The Capillary number compares the effects of viscosity to that due to surface tension and is given by Ca = $\nu \rho \dot{z}/\gamma$, where $\nu$ is the kinematic viscosity and $\dot{z}$ is the droplet's descending speed. 
The maximum speed measured in our experiments is $\dot{z} = 0.025$~m/s for $\nu = 50$~cSt, leading to Ca $= 0.06$. 
This low capillary number suggests that surface tension forces dominate over viscous forces. 
The Weber number, defined as We $= \rho \dot{z}^2 L/\gamma$, compares inertial effects to surface tension with $L$ the characteristic length of the droplet. Taking $L \simeq 10^{- 3}$~m yields a We significantly smaller than unity (We $< 10^{-2}$) across all the experimental parameters tested in this study, indicating the dominance of surface tension over inertial forces. 
Finally, the Reynolds number is given by Re $= \dot{z}L/\nu = 0.5$, which indicates that inertial effects are comparable to viscous effects and may play a role in the droplet's dynamics.
\section{Discussion}

\begin{figure}[ht]
\centering
\centering
\includegraphics[width=\linewidth]{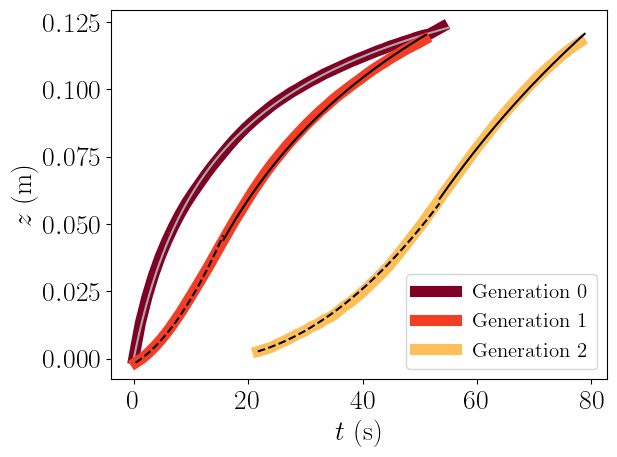}
\caption{\justifying Vertical position $z$ over time $t$ of a droplet sliding along a straight fiber, with different colors representing different generations. The dark red color refers to generation $0$ or mother droplet, the red color represents generation $1$ or the first daughter droplet, and the orange color corresponds to generation $2$, or the granddaughter droplet. The solid lines are fits of Eq. (\ref{Eq_mother_zt}) for generation $0$ and Eq. (\ref{Eq_daughter_zt_shrinking}) for generations $1$ and $2$. The dashed line is the fit of Eq. (\ref{Eq_daughter_zt_growth}). These models effectively describe the different dynamics for all generations. The positions of each droplet tend to converge towards the same value, indicating the merging of the droplets. After this merging event, the refilled mother droplet continues its descent. The experimental conditions for this graph are a droplet volume $\Omega = 5\ \mu$L, viscosity $\nu = 50$~cSt, and fiber diameter $d = 0.35$~mm. }
\label{fig_Mother_zt}
\end{figure}

The location of the droplets from different generations is recorded over time as shown in Fig.~\ref{fig_Mother_zt}, where lighter colors indicate younger generations.
The mother droplet (generation 0) monotonically decelerates, while the subsequent generations behave differently.
The motion of generation 1 and 2 droplets is characterized by an accelerating phase followed by a decelerating phase. 
We begin our analysis by examining the dynamics of the mother droplet.

\subsection{Mother droplet}

The mother droplet, or generation 0, refers to the initial droplet deposited on the dry vertical fiber. 
The dynamics of this droplet have been studied by Gilet~\textit{et al.}~\cite{Gilet2010} and are briefly reviewed in this section. 
The droplet descends the dry fiber driven by its weight $W = \rho g \Omega_{0}$. 
Opposing this driving force is a dissipative force, $F_{\nu} = \xi_{dry} \nu \rho d v $, where $\xi_{dry}$ is a parameter that describes the dissipation of the droplet on a dry fiber.
Balancing these forces gives the speed of the mother droplet
\begin{equation}
    \dot{z}_{0} = \frac{ g}{\xi_{dry} \nu d} \Omega_{0},
    \label{eq_v_gene0}
\end{equation}
where the subscript $[\cdot]_0$ refers to generation $0$. 
As the droplet descends, it loses volume at a rate given by
\begin{equation}
    \dot{\Omega}_{0} = - \pi d \delta_{0} \dot{z}_{0},
    \label{eq_2}
\end{equation}
where $\delta_{0}$ is the thickness of the liquid film left behind the droplet, which depends on the droplet's speed and the fiber diameter. 
This film thickness is given by the Landau-Levich-Derjaguin theory \cite{Landau1988} adapted for a cylindrical fiber as provided in \cite{Quere1999},
\begin{equation}
    \delta_{0} = c_{dry} d \left( \frac{\nu \rho}{\gamma} \dot{z}_{0} \right)^{2/3},
    \label{eq_delta}
\end{equation}
where $c_{dry}$ is a coefficient, with theoretical developments leading to $c_{dry} = 0.67$ \cite{Quere1999}.
Combing \eqref{eq_v_gene0}, \eqref{eq_2} and \eqref{eq_delta} leads to the following differential equation
\begin{equation}
    \ddot{z}_{0} = - w_0 \dot{z}_0^{5/3},
    \label{eq_mother}
\end{equation}
parameterized by 
\begin{equation}
w_0 = \pi \frac{c_{dry}}{\xi_{dry}} \frac{ g}{\nu} d \left(\frac{\nu \rho}{\gamma}\right)^{2/3}.
\label{eq_mother_w0}
\end{equation} 
Equipping initial conditions (i.e., $z_0(t=0) = 0 $ for initial position and $\dot{z}_0(t= 0) = \dot{z}_{0,i}$ for initial speed of droplet generation 0) and solving Eq. \eqref{eq_mother} yields
\begin{equation}
    z_0 = \frac{\dot{z}_{0,i}^{1/3} - \left( \dot{z}_{0,i}^{-2/3} + \frac{2}{3} w_0 t \right)^{-1/2}}{w_0/3}.
    \label{Eq_mother_zt}
\end{equation}
The model accurately depicts the measured dynamics for generation $0$, as shown in Fig.~\ref{fig_Mother_zt} (solid line).

This model is based on the experimental trajectories. 
The parameters $\xi_{dry}$ and $c_{dry}$ are estimated as described in the Supplementary Materials. 
The obtained values are listed in Table \ref{Table_c_xi}.

\begin{table}
 \centering
{\renewcommand{\arraystretch}{2.2}
\setlength{\tabcolsep}{8pt} 
\begin{tabular}{c c c c}
\hline
\hline
  $\xi_{dry}$ & $\xi_{wet}$ & $c_{dry}$ & $c_{wet}$\\
  \hline
   130 & 59 & $1.246 \pm 0.184$ & $0.268 \pm 0.063$\\
   \hline
   \hline
\end{tabular}}
\caption{\justifying Table comparing the values of $\xi$ and $c$ for different fiber states (dry and wet). 
The parameter $\xi$ is estimated using theoretical calculations (see Supplementary Materials, Eqs. (1) and (2)), while the coefficient $c$ is obtained from the fitted prefactor $\xi / c$ and the theoretical estimation of $\xi$ for both dry and wet conditions. 
Both $\xi$ and $c$ are smaller for the wet scenario compared to the dry state. }
\label{Table_c_xi}
\end{table}


\subsection{Daughter droplet}

The daughter droplet forms on the film left by the mother droplet due to Rayleigh-Plateau instability. 
As the liquid film continuously thins down the fiber, the daughter droplet accelerates first before deceleration, as illustrated in Fig.~\ref{fig_Mother_zt}.
The daughter droplet accelerates right after birth as it grows in volume by accumulating liquid left by its progenitor. 
This first stage is called the `growing phase' and can be observed in Fig. \ref{fig_merging} (a). 
Once the inflow at the leading edge balances the outflow at the trailing edge, the droplet reaches its maximal speed. 
Beyond this point, volume loss dominates, and the droplet decelerates. This phase is termed the `shrinking phase'. 
In Fig.~\ref{fig_Mother_zt}, the change in curvature of the droplet's position over time indicates the transition from an accelerating phase to a decelerating one. 
The shrinking phase mirrors the dynamics observed in the mother droplet. 
We propose two models to describe these distinct phases. 

\paragraph{Growing phase}

The forces acting on the generation 1 droplet are similar to those acting on the mother droplet, except that the dissipative coefficient is reduced because the fiber is wet. 
The dissipative coefficient for this generation is written as $\xi_{wet}$. 
The speed of the first daughter droplet is therefore given by
\begin{equation}
    \dot{z}_{1} = \frac{ g}{\xi_{wet} \nu d} \Omega_{1},
    \label{Eq_daughter_v}
\end{equation}
where the subscript $[\cdot]_1$ refers to the first generation. 
As the droplet slides, its volume, $\Omega_{1}$, increases. 
The droplet gains volume from the liquid coating in front of it, while losing a negligible amount at its trailing contact line. 
The rate of volume gained is expressed as
\begin{equation}
\dot{\Omega}_{1,growth} = + \pi d \frac{\bar{\delta_0} z_{tr}}{ t_{tr}},
\label{Eq_daughter_A_dotOmega}
\end{equation}
where $\bar{\delta_0}$ is the average film thickness left by the previous droplet, $z_{tr}$ and $t _{tr}$ are the characteristic length and time over which the droplet grows and correspond to the coordinates of the transition between each phase. 
By combining Eq. (\ref{Eq_daughter_v}) and Eq. (\ref{Eq_daughter_A_dotOmega}), and by considering initial conditions $z_1(t=0) = z_{1,i}$ and $\dot{z}_1(t = 0) = \dot{z}_{1,i}$ we obtain,
\begin{equation}
    z_{1,growth} = A t^2 + \dot{z}_{1,i} t + z_{1,i},
    \label{Eq_daughter_zt_growth}
\end{equation}
with $A$ being a parameter given by
\begin{equation}
    A = \frac{\pi}{2} \frac{ g}{\xi_{wet} \nu} \frac{\bar{\delta_0} z_{tr}}{t_{tr}}.
    \label{Eq_daughter_A}
\end{equation}
To fit the model to the growing phase of the data, $\bar{\delta_0}$ is used as a fitting parameter, while the initial speed $\dot{z}_{1,i}$ is determined by fitting a slope on the first $5\%$ of the data corresponding to the growing phase. 
The position $z_{tr}$ and time $t_{tr}$ are related to the transition between regimes and are extracted from the experimental data, as explained at the end of this section. 
This model captures well the initial part of the droplet's trajectory, as shown by the dotted line in Fig. \ref{fig_Mother_zt}. 
Once the liquid film is no longer thick enough to sustain the increase in volume, the dynamic transitions into a different phase. 

\paragraph{Shrinking phase}
In the second phase, the droplet continues to descend on a wet fiber but decelerates because of its decreasing volume, much like the previous mother droplet. 
The daughter droplet leaves at its rear a liquid film larger than the upcoming film at its front. 
To describe this dynamic, we neglect the input from the front as the output at the rear dominates. 
Therefore, the model is similar to the one of the mother droplet. The rate of volume loss is given by
\begin{equation}
    \dot{\Omega}_{1, shrink} = - \pi d \delta_{1} \dot{z}_{1},
    \label{Eq_daughter_dotOmega}
\end{equation}
where $\delta_1$ is the film thickness left at the rear of the generation 1 droplet. 
This thickness is expressed as
\begin{equation}
    \delta_1 = c_{wet} d \left( \frac{\nu \rho}{\gamma} \dot{z}_1 \right)^{2/3},
    \label{Eq_daughter_delta}
\end{equation}
with $c_{wet}$ being a coefficient. 
To solve the differential equation given by the combination of Eqs. (\ref{Eq_daughter_v}), (\ref{Eq_daughter_dotOmega}) and (\ref{Eq_daughter_delta}), we consider $t_{tr}$ the initial time of the shrinking phase,  $z_1(t = 0) = z_{tr}$ the initial position and $\dot{z}_1 (t = 0) = \dot{z}_{tr}$ the initial speed. 
Actually, $\dot{z}_{tr}$ is the initial speed in this second phase, but is also the maximal speed of the daughter droplet. 
It corresponds to the maximum slope at the transition from the growing phase to the shrinking phase, whose coordinates are $z_{tr}$ and $t_{tr}$. 
We then get this expression for the daughter droplet position in the shrinking phase,
\begin{equation}
    z_{1,shrink} = \frac{ \dot{z}_{tr}^{1/3} - \left( \dot{z}_{tr}^{-2/3} + \frac{2}{3} w_1 \left( t - t_{tr} \right) \right) ^{-1/2}}{w_1/3} + z_{tr},
    \label{Eq_daughter_zt_shrinking}
\end{equation}
with $w_1$ a coefficient given as
\begin{equation}
    w_1 = \pi \frac{c_{wet}}{\xi_{wet}} \frac{ g}{\nu } d \left(\frac{\nu \rho}{\gamma}\right)^{2/3}.
    \label{Eq_daughter_w1}
\end{equation}
This model is fitted to the experimental data in the shrinking phase, as shown in Fig. \ref{fig_Mother_zt} (solid line), with $c_{wet}/\xi_{wet}$ being a fitting parameter. 
The dynamics observed in this phase is similar to the one of the generation~0 droplet.

\begin{figure}[ht]
  \centering
  \begin{subfigure}{\linewidth}
    \centering
\includegraphics[width=\linewidth]{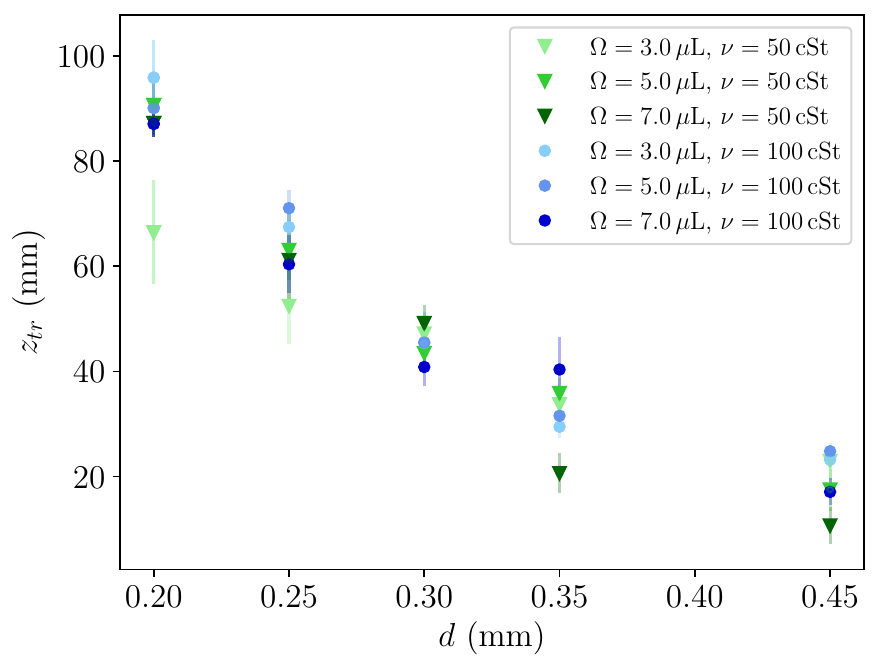}
    \caption{}
  \end{subfigure}

  \vspace{0.5cm} 
  \begin{subfigure}{\linewidth}
    \centering
\includegraphics[width=\linewidth]{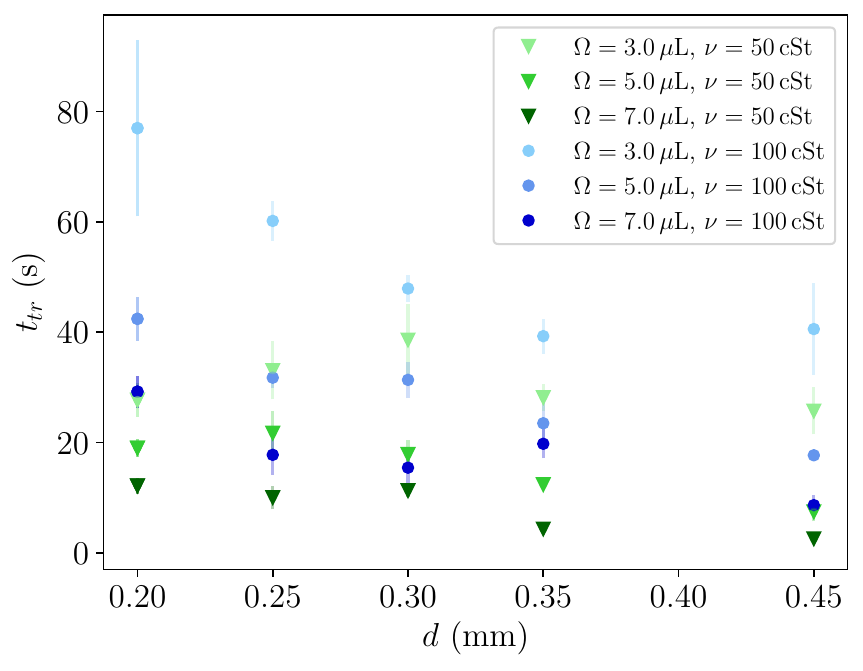}
    \caption{}
  \end{subfigure}
  \caption{\justifying (a) Position of the transition $z_{tr}$ between the growing and the shrinking phase as a function of the diameter $d$ of the fiber. The transition occurs after traveling a smaller distance if the fiber diameter is large. It is independent of both the droplet volume and fluid viscosity. (b) Time after which the transition between the growing and the shrinking phase occurs, $t_{tr}$, as a function of the fiber diameter $d$. The daughter droplet enters the shrinking phase sooner if the fiber diameter is small, the droplet volume is large, and if the fluid is less viscous.  }
  \label{fig_tr}
\end{figure}

The experimental data are divided into two phases to adapt our models. 
It is achieved by identifying the point of maximum slope that separates the accelerating and decelerating portions of the trajectory. 
The position and time at which the transition occurs are $z_{tr}$ and $t_{tr}$, respectively.
They are plotted as a function of the fiber diameter in Fig.~\ref{fig_tr}. 
Both the transition position and time decrease as the diameter increases. 
For a given fiber diameter, the transition seems to occur around the same position, regardless of the droplet volume or viscosity.
The transition time decreases with a large droplet volume and/or small viscosities. 

The obtained values of $\xi_{wet}$ and $c_{wet}$ are summarized in Table \ref{Table_c_xi}. 
Due to the different dynamics of generation 1 compared to generation 0, with smaller coefficients, all generations eventually converge, as shown in Fig.~\ref{fig_merging} (b) and Fig.~\ref{fig_Mother_zt}.

\subsection{Granddaughter droplets}
We have described the dynamics of the mother and the daughter droplets (generations 0 and 1). 
The daughter droplet may also initiate the formation of a granddaughter (generation~2) at the initial position. 
In Fig. \ref{fig_Mother_zt}, the position over time of generation~2 is presented. 
The dynamic of generation~2 is similar to that of generation~1, and the models can also be adapted to the data. 
We anticipate that all subsequent generations will follow the same trend, with the left liquid film being thinner after each subsequent generation, until the thickness is insufficient for the instability to take place and create a droplet. 
As the position of transition from the growing phase to the shrinking phase seems independent of the droplet volume (see Fig.~\ref{fig_tr} (a)), the following generations are expected to transition at that same location. 
In Fig.~\ref{fig_Mother_zt}, the maximal slope is reached at nearly the same position. 
The transition time is expected to increase with each subsequent generation as each new droplet starts with an increasingly smaller initial volume. 


\subsection{Maximum generation}
In this section, we propose a theoretical framework to predict the number of droplet generations based on initial conditions at the mother droplet deposition. 
The two key parameters required for this prediction are the fiber diameter and the volume of the generation 0 droplet. 
The initial liquid film thickness $\delta_{0,i}$ left by the mother droplet near the release point is given by Eq. (\ref{eq_delta}) where $\dot{z}_{0,i}$ is the initial speed. 
The subscript $i$ indicates the initial condition. The liquid film subsequently destabilizes into droplets according to the Rayleigh-Plateau instability, with a characteristic wavelength
\begin{equation}
\lambda = \pi \sqrt{2} d.
\end{equation}
Considering the initial liquid film thickness $\delta_{0,i}$, the volume of the first daughter droplet (generation $1$) can be approximated by
\begin{equation}
 \Omega_{1,i} = \pi^{2} \sqrt{2} d \left( \left( \frac{d}{2} + \delta_{0,i} \right)^{2} - \left(\frac{d}{2}\right)^{2} \right).
\end{equation}
This daughter descends with an initial speed $\dot{z}_{1,i}$ that depends on $\Omega_{1,i}$, and leaves behind a liquid film of thickness $\delta_{1,i}$, given by Eq. (\ref{eq_delta}) with $\dot{z}_{1,i} = g \Omega_{1,i} / \xi_{wet} \nu d $. 
This process may continue, generation 1 droplet may initiate the formation of a subsequent generation (generation $2$), and so on. However, the Rayleigh-Plateau instability stops on vertical fibers when the film thickness $\delta$ becomes smaller than or equal to a critical cutoff value $\delta_c $ given by
\begin{equation}
\delta_c = c_c \frac{ d^{3} }{l_c^{2}} 
\label{eq_deltacut}
\end{equation} where $c_c = 0.175$ and $l_c$ is the capillary length \cite{Quere1990EPL}. 
The last generation will leave a liquid film thickness no thicker than this cutoff value. 
Using this information, we can establish an iterative relationship to calculate the number of generations initiated by the mother droplet.
With $k$ as the iterative parameter, we iterate the following system of equations until $\delta_{k,i} = \delta_c$,

\begin{equation}
    \left\{
    \begin{array}{lll}
        \delta_{k,i} = c d \left( \frac{\nu \rho}{\gamma} \dot{z}_{k,i} \right)^{2/3},  \\
        \Omega_{k+1,i} = \pi^{2} \sqrt{2} d \left( \left( \frac{d}{2} + \delta_{k,i} \right)^{2} - \left( \frac{d}{2} \right)^2 \right),\\
        \dot{z}_{k+1,i} = \frac{ g}{\nu d \xi} \Omega_{k+1,i}.
    \end{array}
\right.
\label{eq_iteration}
\end{equation}
For $k=0$, it refers to the mother droplet, which travels down along a dry fiber. 
For this case, we use the initial speed given by $\dot{z}_{0,i} =  g \Omega_{0,i} / \xi_{dry} \nu  d$. 
For $k \geq 1$, it refers to the $k^{th}$ daughter droplet, which descends along a wet fiber. 
Here, the initial speed is given by $\dot{z}_{k,i} = g \Omega_{k,i} / \xi_{wet} \nu d$. In sections IV. A. and B., we determined the values of $\xi_{dry}$, $c_{dry}$, $\xi_{wet}$, and $c_{wet}$ as summarized in Table \ref{Table_c_xi}. 
The iteration process with parameter $k$ provides the number of generations that a mother droplet will create at the initial position. 
Note that we are analyzing only the initial position of the deposited founding droplet; only the information regarding this initial position is therefore useful, whatever the dynamics and phases downstream. 

\begin{figure}[ht]
  \centering
\includegraphics[width= \linewidth]{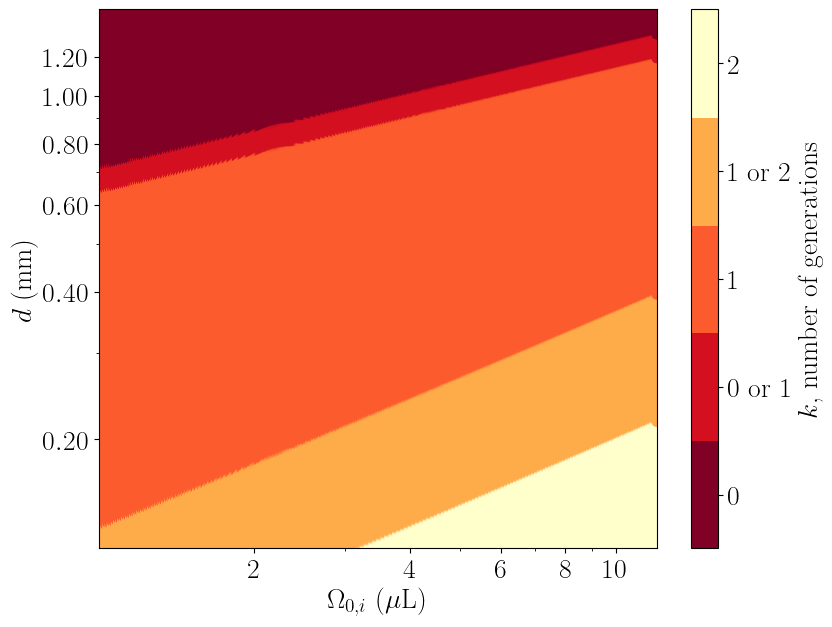} 
\caption{\justifying Number of generations plotted in a double logarithmic graph of the diameter fiber $d$ as a function of the initial mother droplet volume $\Omega_{0,i}$. The different colors represent different generations. The number of generations is calculated by iterations of the system of equations (\ref{eq_iteration}), with the cutoff condition given by Eq. (\ref{eq_deltacut}). The number of generations at the initial deposition point increases with larger droplet volumes and smaller fiber diameters.}
  \label{gene_exp_loglog}
\end{figure}

Figure \ref{gene_exp_loglog} presents the predicted number of generations (indicated by a color scale) in a double logarithmic plot, with the fiber diameter $d$ on the x-axis and the mother droplet volume $\Omega_{0,i}$ on the y-axis. 
The experimental fluctuations lead to varying values of $c_{dry}$ and $c_{wet}$, which affect the transition between $k$ and $k+1$ generations. 
We have thus included buffer areas where uncertainty between $k$ or $k+1$ generations remains. 
The number of generations increases with larger initial droplet volumes and smaller fiber diameters. 
With realistic experimental values as in our studies, for $d$ and $\Omega$, one could observe up to $k=2$ generations, so a daughter followed by a granddaughter. 
Interestingly, for fibers with sub-millimetric diameters, which are commonly used in fiber-based microfluidic devices, always exhibit at least one generation \cite{Weyer2015, Gilet2009, Park2013, Chinju2000}. 
Larger fiber diameters are expected to prevent any droplet generation but experimentally, a microliter droplet on such a large substrate has a clamshell shape rather than a barrel shape as studied in the present analysis \cite{McHale2002}.

Experimentally, it is challenging to measure the exact number of generation droplets at the deposition point. 
The droplet is released using a syringe, but since it is manually handled, slight variations can occur when transferring the droplet onto the fiber. 
As a result, the liquid film at that position is influenced not only by the sliding of the initial droplet but also by the contact with the syringe or by the droplet pinching. 
Additionally, an unstable clamshell shape may initially appear before transitioning into the stable barrel shape. 
Although experiments with observable clamshell shapes are excluded, this shape can be metastable and briefly appear within the first millimeter of descent \cite{Kern2024}. 
Another limitation is the small size of the subsequent generation droplets at the release point, making them difficult to detect accurately. 
Furthermore, thin liquid films destabilize over extended timescales, sometimes requiring hours to break up into droplets. 
Despite these challenges, at least one generation has been observed in all experiments. 

The system of equations in Eqs. (\ref{eq_iteration}) can also be reformulated as a nonlinear recursive sequence, which yields 
\begin{equation}
    \delta_{k+1} = \alpha d \left( \delta_{k}^2 + d \delta_{k}\right)^{2/3},
\end{equation}
where $\alpha$ is a coefficient given by 
\begin{equation}
    \alpha = c_{wet} \left( \frac{\sqrt{2} \pi^2}{\xi_{wet}} \frac{\rho g}{\gamma}  \right)^{2/3}.
\end{equation}
The system has three fixed points, $\delta_0$, $\delta_-$ and $\delta_+$ given by
\begin{equation}
    \delta_0 = 0,
\end{equation}
\begin{equation}
    \delta_- = \frac{1}{2} \left( \frac{1}{\alpha^3 d^3} - 2d - \frac{\sqrt{1-4 \alpha^3 d^4}}{\alpha^3 d^3} \right),
    \label{Eq_deltaMinus}
\end{equation}
and
\begin{equation}
    \delta_+ = \frac{1}{2} \left( \frac{1}{\alpha^3 d^3} - 2d + \frac{\sqrt{1-4 \alpha^3 d^4}}{\alpha^3 d^3} \right).
\end{equation}
Other than the trivial solution $\delta_0$,  $\delta_+$ is an unstable fixed point, and the sequence $\delta_k$ converges towards $\delta_{-}$. The analysis of the fixed points and their stability is detailed in the Supplementary Materials.

\begin{figure}[ht]
\centering
\centering
\includegraphics[width=0.48 \textwidth]{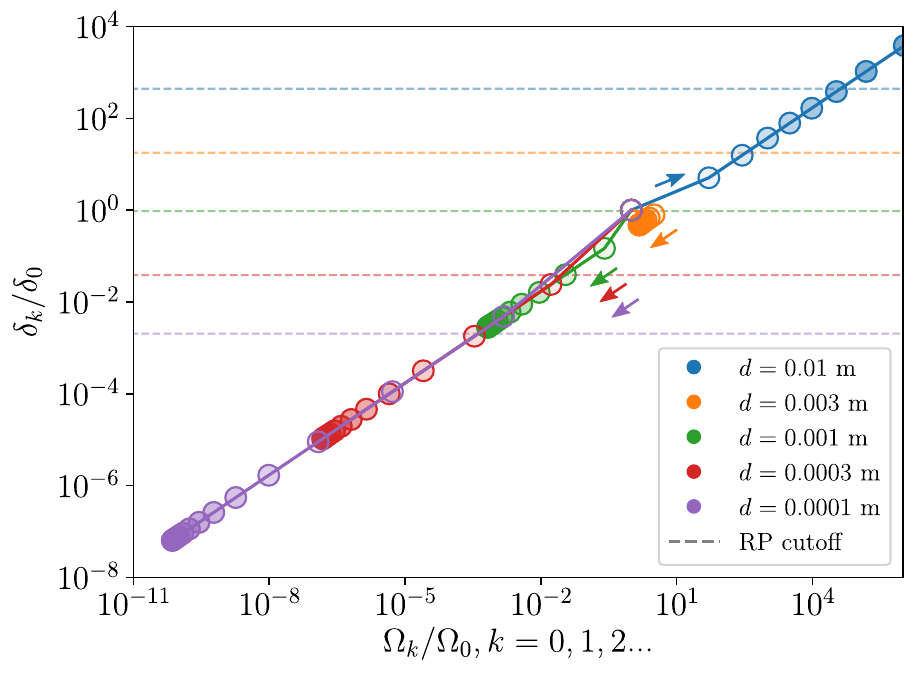} 
\caption{\justifying Iterative values of the dimensionless film thickness $\delta_k/\delta_0$ as a function the iterative values of the dimensionless droplet volume $\Omega_k/\Omega_0$ for different fiber diameters $d$. Darker markers indicated later iterations. An interesting phenomenon emerges, for sufficiently small fiber diameters, the sequence converges to a fixed point, while for larger diameters, it diverges. The dotted horizontal lines represent the Rayleigh-Plateau cutoff, as defined in Eq. (\ref{eq_deltacut}). The arrows show the direction of iteration.}
\label{fig_iteration}
\end{figure}

The condition for convergence, $1-4 \alpha^3 d^4 >0$, imposes an upper bound on the fiber diameter,
\begin{equation}
    d < \left( 2 \sqrt{2} \pi^2 \frac{c_{wet}^{3/2}}{\xi_{wet}} \frac{\rho g}{\gamma} \right)^{-1/2}.
    \label{d_max_FixedPoints}
\end{equation}
For the numerical value used in this study, the critical fiber diameter is $d < 5.7$~mm. 
Notably, this value approximates the droplet diameter, $2 \Omega^{1/3} \approx 3.4$~mm. This suggests a physical interpretation: beyond this threshold, the fiber is too large to sustain a stable barrel shape droplet compatible with the assumptions of our model. 
For large fibers, the droplet adopts a clamshell shape, which alters the dynamics due to the droplet asymmetry, i.e., reduced contact area with the fiber and asymmetric liquid film left on the fiber.
Furthermore, for large fiber diameter, the largest droplet radius that can be supported on a horizontal fiber tends to a constant value of $1.6 l_c$~\cite{Lorenceau2004JCol}. 
In such a case, one obtains a maximal Bond number of
\begin{equation}
    \mathrm{Bo_{up}} = \frac{\rho g (1.6 l_c)^2}{\gamma} = 1.6^2 = 2.56
\end{equation}
that falls outside the applicability of our model. 
Here the fiber has negligible influence on the droplet, with almost no dissipation at the contact line or liquid wedge.
The existence of the fixed points, given by Eq. (\ref{d_max_FixedPoints}), seems to be linked to the domain of validity of our model (Bo $\sim 1$).
The convergence for small $d$ and the divergence for large $d$ are observed in Fig. \ref{fig_iteration}. 
This figure shows the evolution of the film thickness $\delta_k$ as a function of the iterated droplet volume $\Omega_k$ for various fiber diameters $d$. 
The initial volume corresponds to the experimental value, i.e. $\Omega_{0,i} = 5~\mu$L. 
Darker circles indicate later iterations (i.e., increasing $k$). 
For $d$ larger than the critical fiber diameter, the sequence diverges with iteratively increasing volume and coating, while for smaller $d$, both quantities progressively shrink toward a limiting value.
Choosing appropriate values for $\Omega_{0,i}$ and $d$ ensures a final nonzero $\delta_-$ for the film coating. 
The value of this fix point remains always smaller than $\delta_c$ for all $d$, thus the mathematical convergence to $\delta_-$ cannot be observed experimentally. 
In all cases, the droplet birth cycle of droplets vanishes in both the mathematical model and the experiment, although the mechanisms and the end differ in each case.

\subsection{Other families}

We have analyzed the number of generations at the initial deposition location of the mother droplet. 
Are there other potential locations where other kinds of families could appear? 
A droplet may appear if the coating is thicker in a given place. 
This is the case at the merging position between subsequent generations, as for the daughter droplet meeting with the mother droplet observed in Fig. \ref{fig_merging} (b). 
The reunited droplets accelerate together due to the merging, creating a thicker liquid film, as observed and analyzed in \cite{Lee2021, Feng2024}. 
As the film thickness is larger at that particular location, the film may destabilize into a droplet and initiate a new family. 
Furthermore, all droplets descending the non-uniformly coated fiber reach a maximal speed between their growing and shrinking phases. 
Larger speed induces larger film thickness. 
The position of the transition could also be the birth position of another family. 
These cases are not analyzed further in the present paper as these families are more difficult to predict and complex to observe experimentally.

\section{Conclusion}

In this paper, we study the motion of a droplet along a vertical fiber and the residual liquid film that coats the fiber as the droplet descends. 
The speed of the droplet controls the coating thickness. 
Thus, the initial droplet, which slows down as it descends, leaves a non-uniform film, thickest film the release point. 
The coating film destabilizes through the Rayleigh-Plateau instability, forming droplets faster where the liquid film is thicker, i.e. the release point of the droplet. 
This newly formed droplet grows by accumulating liquid from the film ahead of it as it slides down. 
The initial droplet is referred to as generation 0 (mother), while the droplet resulting from the liquid destabilization is termed generation 1 (daughter). 
The dynamics of generations 0 and 1 are distinct, with generation 0 sliding on a dry fiber and generation 1 descending on a wet fiber.
This difference leads to the merging of both generations after traveling a certain distance. Generation 1 also leaves behind a liquid film, potentially leading to further droplet formation and continuing the cascading cycle.
``It's as if time had turned around and we were back at the beginning"---quote from Márquez~\cite{GarciaMarquez}. 
We provide a theoretical estimation of the number of generations created at the release point of the initial droplet. 

This study contributes to the extensive literature on droplet behavior on fibers. 
It serves as a further step towards understanding the dynamics of self-sustaining droplets on fibers and opens up several avenues for further research. 
For instance, what occurs after the merging of the mother and daughter droplets? 
The ``re-fed" mother droplet may initiate a new cycle, and analyzing the long-term behavior, including potential oscillations and increases in period, could provide interesting insights.
These results are relevant for fiber coatings and fiber-based microfluidics, where generation control is essential.
We show that droplets on fibers can travel further than what is expected, thanks to the cascade of merging. 
This result is important for fog collection on harp fiber arrays.
Beyond fibers, a droplet running down a partially wetting inclined plate depicts a corner tail at its rear that can break into smaller droplets \cite{Podgorski2001, LeGrand2005}.
In such a case 
one could observe similar behavior as reported in this paper.

Our results show how a single droplet initiates a cascade of generations, each leaving traces for the next, much like the Buendía family in One Hundred Years of Solitude, where no generation exists in isolation.
Just as the town of Macondo confines the Buendía family within a closed space, the fiber constrains the motion of the droplets.
Both families and droplets inherit an inescapable fate, whether through familly history and solitude or through a few physical laws such as Plateau–Rayleigh instability and Landau–Levich flows. 
This interplay of forces shapes each generation, producing a rhythm of rise and fall, of birth and merging, until the cycle reaches its inevitable end.
\vskip 0.2 cm
\section*{Acknowledgments} 

This work is financially supported by the University of Li\`ege through the CESAM Research Unit.

\vskip 0.2 cm
\section*{Author contributions} 
J.V.H. conducted the experiments, analyzed the results and took the lead in writing the article. N.V. and Z.P. supervised the project and secured funding. All authors participated in editing the manuscript.

\section*{Note}
The authors declare no competing financial interest.


\bibliographystyle{achemso.bst}
\bibliography{biblio} 

\end{document}